\documentclass[pra,twocolumn,showpacs]{revtex4}

\usepackage{amsmath}
\usepackage{amssymb}
\usepackage{amsthm}
\usepackage{graphicx}

\newcommand{\tr}{\operatorname{tr}}
\newcommand{\uinvnorm}{|\kern-1pt|\kern-1pt|}

\newcommand{\spectrum}{\operatorname{sp}}

\newcommand{\supp}{\operatorname{supp}}

\theoremstyle{plain}

\theoremstyle{definition}

\theoremstyle{remark}

\begin{document}
\bibliographystyle{apsrev}

\title{The ground state of a class of noncritical 1D quantum spin systems can be approximated efficiently}

\author{Tobias J.\ Osborne}
\email[]{Tobias.Osborne@rhul.ac.uk} \affiliation{Department of
Mathematics, Royal Holloway University of London, Egham, Surrey TW20
0EX, United Kingdom}

\date{\today}

\begin{abstract}
We study families $H_n$ of $1$D quantum spin systems, where $n$ is
the number of spins, which have a spectral gap $\Delta E$ between
the ground-state and first-excited state energy that scales,
asymptotically, as a constant in $n$. We show that if the ground
state $|\Omega_m\rangle$ of the hamiltonian $H_m$ on $m$ spins,
where $m$ is an $O(1)$ constant, is \emph{locally} the same as the
ground state $|\Omega_n\rangle$, for arbitrarily large $n$, then an
arbitrarily good approximation to the ground state of $H_n$ can be
stored efficiently for all $n$. We formulate a conjecture that, if
true, would imply our result applies to all noncritical $1$D spin
systems. We also include an appendix on quasi-adiabatic evolutions.
\end{abstract}

\pacs{75.10.Pq, 03.67.-a, 75.40.Mg}

\maketitle

\section{Introduction}

The physics of low-dimensional lattices of quantum spins is
especially rich and varied. As a consequence, a great deal of effort
has gone into understanding the statics and dynamics of these
systems. However, despite this effort, many fundamental questions
about quantum lattice models remain unanswered. Perhaps one of the
most pressing of these questions is: can we \emph{faithfully}
approximate, at least via some efficient numerical procedure, the
ground-state properties of quantum lattice systems? If this were not
true for physically realistic models then we would have to give up
any hope of extracting theoretical predictions from these models.

It is a folk theorem that, at least in one dimension, an
approximation to the ground-state properties of a \emph{noncritical}
\cite{endnote28} chain of quantum spins may be obtained efficiently
on a classical computer. Significant progress towards proving this
theorem has been obtained recently in a ground-breaking paper by
Hastings \cite{hastings:2006a}. Hastings found a procedure whereby
an approximation to the ground state of a noncritical quantum spin
system could be obtained and stored using subexponential resources
that scale as $n^{c\log(n)}$, where $c$ is some constant which
depends on the spectral gap $\Delta E$ and the local spin dimension.
The computational complexity of this method is not far off the
expected result, i.e., $n^r$, where $r$ is some constant.

There is one procedure which appears to provide arbitrarily good
approximations to the ground-state properties of $1$D noncritical
quantum spin systems, namely, the density matrix renormalisation
group (DMRG). (See \cite{schollwoeck:2005a} and references therein
for a description of the DMRG and its relatives.) The DMRG is the
premier tool used in numerical explorations of the physics of 1D
quantum systems, and has been used with unparalleled success in
simulating both their statics, and more recently, dynamics. Many
exciting extensions of the DMRG have been developed, including, a
powerful variant for 2D systems \cite{verstraete:2004a}.

Unfortunately the DMRG is not known to be \emph{correct}. That is,
it is unclear if the DMRG always faithfully returns an approximation
to the ground state and not some other low-lying excited state.
Additionally, the \emph{complexity} of the DMRG is currently
unknown. It is entirely possible, in principle, that the DMRG
requires exponential resources to obtain a faithful approximation to
the ground-state of noncritical spin system. However, extensive
numerical experimentation strongly indicates that the DMRG requires
only \emph{linear} resources in $n$ to simulate noncritical systems.
Nevertheless, hard instances for variants of the DMRG do exist
\cite{eisert:2006a}, which means that we must be cautious when
applying the DMRG in certain situations. This strongly suggests that
while the \emph{average-case} complexity of the DMRG may be
polynomial, the \emph{worst-case} complexity is probably
exponential.

The DMRG can be thought of as a variational minimisation of the
energy over the class of \emph{finitely correlated states} or
\emph{matrix product states} (for an introduction to finitely
correlated states and a detailed description of their properties see
\cite{fannes:1992a}). Finitely correlated states are particularly
well-suited to this task because: (a) it is expected that they
approximate realistic ground states well; and (b) there is an
efficient computational procedure to extract local properties, like
correlators, from a state stored as a finitely correlated state. The
validity of the FCS ground-state ansatz is conditioned, at least, on
the truth of (a), thus it is very desirable to show that good
approximations to the ground states of some physically interesting
class of spin systems could be stored efficiently as a FCS.

There are many ways to obtain an approximation to the ground state
of a quantum system. For example, in the case of the DMRG, there are
variants \cite{vidal:2003a, zwolak:2004a, verstraete:2004b} which
obtain ground-state approximations via imaginary time evolution.
However, in this paper, we'd like to emphasise another method to
obtain ground-state approximations, namely, via adiabatic
continuation. The idea with adiabatic continuation is to start with
a hamiltonian $H(0)$ whose ground-state is known exactly, and then
to adiabatically vary along a path of hamiltonians $H(s)$ until the
desired hamiltonian $H(1)$ is reached. Under the adiabatic dynamics,
the state at the end of the evolution will be the ground state of
the final hamiltonian $H(1)$.

The spectral gap $\Delta E(s)$ between the ground- and first-excited
state energy of $H(s)$ provides the fundamental obstruction to
approximating adiabatic dynamics: the smaller $\Delta E(s)$ is, the
harder it is to approximate the dynamics. At this point we'd like to
point out an obvious (but crucial) fact: if arbitrary paths $H(s)$
are allowed then $\Delta E(s)$ can be made as large as desired up to
$\min\{\Delta E(0), \Delta E(1)\}$. However, in the context of spin
systems, we don't allow arbitrary paths because they would
presumably lead to an unphysical situation where $H(s)$ contains
interactions between many spins. Rather, we demand that $H(s)$ has
only \emph{local} interactions throughout the path $s \in [0, 1]$.
This additional constraint motivates us to define the notion of
\emph{adiabatic connectivity}: two quantum spin systems $H$ and $K$
are said to be \emph{adiabatically connected} if there exists a path
of \emph{local} hamiltonians $L(s)$ such that $L(0) = H$ and $L(1) =
K$, and the spectral gap for $L(s)$ satisfies $\Delta E(s) >0$ for
all $s\in[0,1]$.

In the case that a hamiltonian $H$ is adiabatically connected to
another hamiltonian $K$ via a path $L(s)$ with $\Delta E(s) >
\text{const.}$ it turns out that ground-state properties of $H$ can
be \emph{efficiently} and \emph{certifiably} obtained from those of
$K$ \cite{osborne:2006a}. Because of certain counterexample systems
for DMRG methods \cite{endnote29} it appears that adiabatic
continuation is the only method whereby certifiable approximations
to the ground-state can be obtained efficiently. Thus, the problem
of understanding the ground-state properties of a quantum spin
system $H$ can be reduced to finding a hamiltonian for a
well-understood spin system $K$ which is adiabatically connected to
$H$.

In this paper we consider the problem of proving that the isolated
eigenstates of a certain class of noncritical quantum spin systems
can be efficiently represented as finitely correlated states with
polynomial computational storage resources (in $n$ and $1/\Delta
E$). (The reason we say ``isolated eigenstates" here is because our
subsequent derivations make no use of the fact that the eigenstate
in question is the ground state. For example, the argument applies
equally to the highest-energy eigenstate.) The noncritical systems
we consider are local hamiltonians which satisfy a crucial
additional requirement: we assume that the ground state
$|\Omega_m\rangle$ of the system on $m$ spins is \emph{locally}
close to the ground state $|\Omega_n\rangle$ for $n$ spins, where
$n$ is arbitrarily large. This is a fundamental physical assumption
which, philosophically, underlies the success of the DMRG and
relatives. In the case that this requirement is satisfied we show
$H_n$ \emph{is} adiabatically connected to a hamiltonian $K$ whose
ground state is exactly and efficiently known. As a consequence, if
we know the ground-state energy $\Omega_n$ for all $n$, we show that
the ground state of $H_n$ may be efficiently approximated by a
finitely correlated state.

\section{Formulation}

We will, for the sake of clarity, describe our results mainly for a
finite chain $\mathcal{C}$ of $n$ distinguishable spin-$1/2$
particles. The family $H$ of local hamiltonians we focus on (which
implicitly depends on $n$) is defined by $H = \sum_{j=0}^{n-2} h_j$,
where $h_j$ acts nontrivially only on spins $j$ and $j+1$. We set
the energy scale by assuming that $\|h_j\|$ scales as a constant
with $n$ for all $j = 0,1, \ldots, n-1$, where $\|\cdot\|$ denotes
the operator norm. We can easily accommodate next-nearest neighbour
interactions etc.\ by blocking sites and thinking of the blocks as
new (larger) spins. However this can only be done a constant number
of times: the quality of our approximation will decrease
exponentially with the number of such blockings. We do not assume
translational invariance.

We make three major assumptions about our system. The first is that
the spectrum of $H= \sum_{j=0}^{2^n-1} E_j |E_j\rangle \langle E_j|$
has a spectral gap $\Delta E = E_1-E_0$ between the ground-state
energy and the first-excited state energy which is always strictly
positive and scales as a constant with $n$. (This is the
\emph{noncriticality} assumption.) The second assumption we make is
that the ground-state energy is set to zero. This is the principle
reason why our analysis does not allow us to, in principle,
efficiently calculate approximations to ground-state properties of
$H$ because to perform this operation we need to know $E_0$ --- it
is potentially a computationally difficult task to approximate the
ground-state energy eigenvalue \cite{oliveira:2005a, kempe:2004a,
kitaev:2002a}. Our final assumption is that the ground state
$|\Omega_m\rangle$ for $m$ spins, with $m$ an $O(1)$ constant, is
\emph{locally similar} to $|\Omega_n\rangle$, with $n$ arbitrarily
large. This means that there exist unitary operators $U$ and $V$
which act nontrivially only on a contiguous block $\Lambda_1$
(respectively, $\Lambda_2$) of a constant number $l\ll m$ of spins
located at the left (respectively, right) end of the block of $m$
spins such that fidelity
\begin{equation}\label{eq:overlap}
x = \langle \Omega_m|U^\dag\otimes \mathbb{I} \otimes V^\dag
\rho^{(n)}_{m} U\otimes \mathbb{I} \otimes V |\Omega_m\rangle
\end{equation}
is an $O(1)$ constant independent of $n$, where $\rho^{(n)}_{m} =
\tr_{\widehat{m}}(|\Omega_n\rangle \langle \Omega_n|)$ is the
reduced density operator for $|\Omega_n\rangle$ on $m$ contiguous
spins. In words: we assume that the ground state $|\Omega_m\rangle$
of $H_m$ has some overlap with $|\Omega_n\rangle$ when we are
allowed to apply some correction operations to the ends of the chain
of $m$ spins. The physical idea underlying this assumption is that
for noncritical spin systems the ground state of $m$ spins ought to
be the same as that for $n > m$ spins apart from boundary effects
which should persist only a distance $l = c/\Delta E$, with $c$ some
constant, into the bulk of the ground state of both systems. While
this is physically reasonable we've been unable to show that it's
true for all noncritical spin systems. Hence we make this an
assumption.

We are going to make a further simplifying assumption about the
systems we are considering, namely that the overlap $x' = \langle
\Omega_m|\rho^{(n)}_{m} |\Omega_m\rangle$ is an $O(1)$ constant.
This is obviously a far stronger assumption than that of the
Eq.~(\ref{eq:overlap}).  However, it turns out that this assumption
entails no loss of generality in our subsequent derivations. The way
to see this is to first notice that the state $U\otimes \mathbb{I}
\otimes V |\Omega_m\rangle$, where $U$ and $V$ are chosen as in
Eq.~(\ref{eq:overlap}), is the unique gapped ground state of $H_m' =
U\otimes \mathbb{I} \otimes V H_m U^\dag\otimes \mathbb{I} \otimes
V^\dag$. As long as $U$ and $V$ act on only a small number of spins
(in comparison to $m$) near the boundary then $H_m'$ will also be a
local hamiltonian. Our subsequent analysis only requires that our
start hamiltonian $H_m'$ contains local interactions.

Before we end this section we introduce some notation for
approximations. If we have two quantities $A$ and $B$ then we use
the notation $A\lesssim B$ to denote the estimate $A\le CB$ for some
constant $C$ independent of $n$. Because we'll be interested in the
consequences of allowing the minimum gap $\Delta E$ and the overlap
$x$ to depend on $n$ we'll explicitly retain any dependence on
$\Delta E$ and $x$ in our calculations.

\section{Adiabatic connections between $H_m$ and $H_n$}

\begin{figure}
\begin{center}
\includegraphics{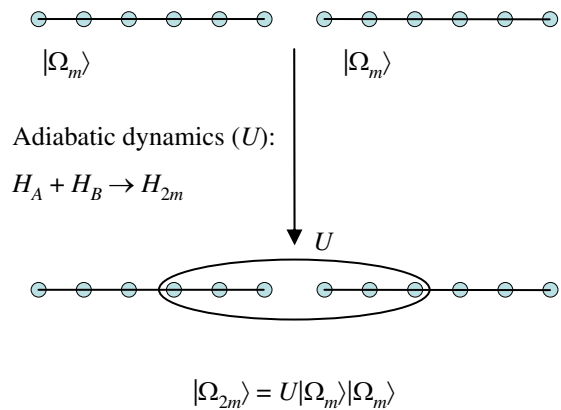}
\caption{Illustration of the procedure to glue two copies of the
ground state $|\Omega_m\rangle$ for $H_m$ together via adiabatic
continuation to obtain an approximation to the ground state of
$H_{2m}$.}\label{fig:gluing}
\end{center}
\end{figure}

In this section we construct a hamiltonian $K$, whose ground state
is known exactly, and which is adiabatically connected to $H_n$. We
construct $K$ iteratively: we show that the hamiltonian $K_m$, which
we define to consist of two copies $A$ and $B$ of $H_m$, is
adiabatically connected to $H_{2m}$. Thus we ``glue'' the ground
states of $A$ and $B$ together via adiabatic continuation. We then
show that the adiabatic continuation from $K_m$ to $H_{2m}$ can be
approximated by a unitary operator which acts on only a constant
number of sites across the boundary between $A$ and $B$. We then
iterate this gluing procedure to obtain the ground state of $H_n$.
See Fig.~\ref{fig:gluing} for a schematic illustration of our
procedure.

We describe our method in a slightly more general context: we show
how to glue together two systems $A$ and $B$ of $m$ and $n-m$ spins,
respectively. The way we show that a good approximation
$|\widetilde{\Omega}\rangle$ to the ground state $|\Omega\rangle$
can be stored efficiently is to consider the (adiabatic) dynamics of
an auxiliary system $\mathcal{K}$ which is constructed in the
following way. First fix $n$, the total number of spins. Next
partition the chain into two contiguous blocks $A$ and $B$. Now
consider the following hamiltonian
\begin{equation}
K = H - H_I
\end{equation}
where $H_I$ is an interaction term, to be defined, which spans the
boundary between the two blocks. ($H_I$ will not be the same as
$h_I$, the interaction term in $H$ which spans the boundary.) If $A$
consists of the first $m$ spins, $m < n$, then $B$ consists of the
last $n-m$ spins and we define $H_I= h_I + \delta\mathbb{I}$, where
$\delta$ is some constant. Thus we can write $K = H_A + H_B -
\delta\mathbb{I}$, where $H_A=\sum_{j=0}^{m-2} h_j$ and $H_B =
\sum_{j=m}^{n-2} h_j$. We note that $K$ has a unique ground state
and spectral gap $\Delta E$ because of the assumed spectral
structure of the \emph{family} $H_n$, i.e., both $H_A$ and $H_B$
belong to the family $H_n$ ($H_A \approx H_m$ and $H_B\approx
H_{n-m}$). We set the constant $\delta$ so that the ground-state
energy of $K$ is zero. (Recall that we've already set the zero of
energy by requiring the ground-state energy of $H$ is $0$.)

Now we construct a new system $\mathcal{C}'$ whose hilbert space is
a \emph{direct sum} of two copies of the old hilbert space. The
hamiltonian $L$ for the new system is a direct sum of $H$ and $K$:
\begin{equation}
L = \left(\begin{matrix}H & 0\\0 & K\end{matrix}\right).
\end{equation}
The hilbert space for our new system is thus given by
$\mathcal{H}_{\mathcal{C}'} = \bigotimes_{j=0}^{n} \mathbb{C}^2$. We
think of this hilbert space as that of the original chain
$\mathcal{C}$ of $n$ spins with an extra spin, which we call $C'$,
that lives between spins $m-1$ and $m$. Thus we can write $L$ as $L
= \mathbb{I}_{C'}\otimes K + \left(\frac{\mathbb{I}_{C'} +
\sigma^z_{C'}}{2}\right)\otimes H_I$, where $\sigma^z =
(\begin{smallmatrix} 1 & 0 \\ 0 & -1\end{smallmatrix})$.

We next observe, by the assumed properties of $H$ and $K$, that the
spectrum $\spectrum(L)$ of $L$ has the following structure. Firstly,
the hamiltonian $L$ has a doubly degenerate ground eigenspace
spanned by the vectors $\{|0\rangle|\Omega_H\rangle,
|1\rangle|\Omega_K\rangle\}$, where $|\Omega_H\rangle$
(respectively, $|\Omega_K\rangle$) is the ground state of $H$
(respectively, $K$). The hamiltonian $L$ then has a gap $\Delta E$
which is larger than some constant, irrespective of the number $n$
of spins.

To construct our final auxiliary system $\mathcal{K}$ we consider
the parameter-dependent hamiltonian
\begin{equation}
M(\theta) = L + \kappa V_{C'}(\theta)\otimes
\mathbb{I}_{\mathcal{C}}
\end{equation}
where
\begin{equation}
V(\theta) = (\sin(\theta)|0\rangle -
\cos(\theta)|1\rangle)(\sin(\theta)\langle0|
-\cos(\theta)\langle1|).
\end{equation}
We note that $V(\theta)$ is positive semidefinite and it acts
nontrivially only on the auxiliary spin. Thus, to find the ground
state of $M(\theta)$ we can restrict our attention to subspace
spanned by $\{|0\rangle|\Omega_H\rangle,
|1\rangle|\Omega_K\rangle\}$.

The addition of the operator $V(\theta)$ will perturb all of the
eigenvectors of $L$. However, by Weyl's perturbation Theorem
\cite{bhatia:1997a}, as long as $\kappa < \frac{\Delta E}{2}$ the
operator $V(\theta)$ will not mix the ground subspace of $L$ with
the remaining eigenvectors of $M(\theta)$; this subspace will always
be separated by a gap from the rest of the spectrum. Furthermore,
the subspace itself is unchanged: only the eigenvectors within this
subspace change under the addition of $V(\theta)$. Thus we fix
$\kappa = \frac{\Delta E}{4}$, so $M(\theta) = L + \frac{\Delta
E}{4} V(\theta)$. The matrix elements of $V(\theta)$ in the ground
eigenspace of $L$ are given by $V_{jk}(\theta) =\langle \psi_j
|V(\theta)|\psi_k\rangle$, $j,k\in\{0,1\}$, where $|\psi_0\rangle =
|0\rangle |\Omega_H\rangle$ and $|\psi_1\rangle = |1\rangle
|\Omega_K\rangle$:
\begin{equation}
V_{jk}(\theta) = \begin{pmatrix}\sin^2(\theta) &
-\sin(\theta)\cos(\theta) \overline{x} \\ -\sin(\theta)\cos(\theta)
{x} & \cos^2(\theta)\end{pmatrix}.
\end{equation}
The two eigenvalues and eigenvectors of this matrix correspond to
the ground state and first excited state of $M(\theta)$. The
corresponding gap of $V(\theta)$ is
\begin{equation}
\Delta(\theta) = \sqrt{1-4\sin^2(\theta)\cos^2(\theta) (1-|x|^2)},
\end{equation}
which has a minimum value equal to $|x|$ at $\theta=\frac{\pi}{4}$,
where $x$ is defined by Eq.~(\ref{eq:overlap}).

We think of the $V(\theta)$ contribution in $M(\theta)$ as
\emph{polarising} the system $L$ so it has a unique gapped ground
state
\begin{equation}
|\Omega_M(\theta)\rangle= \cos(\theta)|0\rangle|\Omega_H\rangle +
\sin(\theta) |1\rangle|\Omega_K\rangle.
\end{equation}
By the discussion in the previous paragraph the gap above this
ground state is always larger than $\frac{\Delta E |x|}{4}$.

The idea behind our proof is now simple to state. We begin with the
system $\mathcal{K}$ in the ground state $|\Omega_M(0)\rangle =
|0\rangle_{C'}|\Omega_A\rangle|\Omega_B\rangle$ of $M(0)$ and
adiabatically vary $\theta$ from $0$ to $\frac{\pi}{2}$. The
resulting ground state is $|1\rangle_{C'} |\Omega_H\rangle$. This is
a product state between the original chain $\mathcal{C}$ and $C'$.
We then discard the ancilla spin $C'$ to obtain the ground state of
$\mathcal{C}$. We approximate this exact adiabatic evolution with a
unitary operator which acts nontrivially only on a small set
$\Lambda$ of spins across the boundary between $A$ and $B$.

\section{Approximating the adiabatic dynamics}

The adiabatic evolution of the ground state
$|\Omega_M(\theta)\rangle = \mathcal{U}(\theta;
0)|\Omega_M(0)\rangle$ of $M(\theta)$ is generated by the solution
of the differential equation
\begin{equation}
\frac{d}{d\theta}\mathcal{U}(\theta; 0) = [\Omega_M'(\theta),
\Omega_M(\theta)]\mathcal{U}(\theta; 0),
\end{equation}
where $\Omega_M(\theta) = |\Omega_M(\theta)\rangle \langle
\Omega_M(\theta) |$. This is an example of an exact adiabatic
evolution (see App.~\ref{app:appendix1} for more discussion of exact
adiabatic evolution).

We approximate the exact adiabatic evolution $\mathcal{U}(\theta)$
by \emph{quasi-adiabatic} evolution which, for us, is defined by the
solution of the differential equation
\begin{multline}
\frac{d}{ds}\mathcal{V}(s;0) = \\ i\int_{-\infty}^{\infty}
\chi_{\gamma}(t)\left(\int_0^t \tau_u^{M(s)}\left(\frac{\partial
M(s)}{\partial s}\right) du\right) dt \,\mathcal{V}(s;0),
\end{multline}
where
\begin{equation}
\chi_\gamma(t) =
\frac{e^{-\frac{t^2}{2\gamma^2}}}{\sqrt{2\pi}\gamma},
\end{equation}
and $\gamma$ is some constant to be set later, and $\tau_t^{M}(N) =
e^{itM}Ne^{-itM}$. (See App.~\ref{app:appendix1} for a discussion of
quasi-adiabatic evolution and related adiabatic-like evolutions.)

The infinitesimal generator
\begin{equation}
k(s) = \int_{-\infty}^{\infty} \chi_{\gamma}(t)\left(\int_0^t
\tau_u^{M(s)}\left(\frac{\partial M(s)}{\partial s}\right) du\right)
dt
\end{equation}
of quasi-adiabatic evolution is an operator which is
\emph{approximately local} in a region $\Lambda_j = \{l \, | \,
d(l,C') \le j\}$ of $2j+1$ sites surrounding the boundary between
$A$ and $B$. The way to see this intuitively is to recognise that
$\frac{\partial M(s)}{\partial s}$ is  strictly local in the region
$\Lambda_1$, and the operator $\tau_u^{M(s)}\left(\frac{\partial
M(s)}{\partial s}\right)$ is a local operator evolved according to a
local hamiltonian for some time which is approximately less than
some constant $\gamma$. We make this intuition precise by applying a
Lieb-Robinson bound \cite{lieb:1972a, hastings:2004a,
nachtergaele:2005a, hastings:2005b} (see \cite{osborne:2006b} for a
simple direct proof). The Lieb-Robinson bound reads (for a system
with hamiltonian $H$)
\begin{equation}
\|[\tau_{t}^{H}(A), B]\| \le |Y| e^{-v d(x,Y)}(e^{\kappa |t|}-1),
\end{equation}
for any two norm-$1$ operators $A$ acting on site $x$ and $B$ acting
on a subset ${Y}$ of sites, with $\{x\}\cap Y = \emptyset$, which
are separated by a distance $d(x,Y)$. The constants $v$ and $\kappa$
are independent of $n$ and depend only on $\|h\|$, which is an
$O(1)$ constant.

What we do is define
\begin{equation}
k_{\alpha}(s) =
\mathcal{F}_s^{M_{\Lambda_{\alpha}}(s)}\left(\frac{\partial
M(s)}{\partial s}\right),
\end{equation}
where
\begin{equation}
\mathcal{F}_s^{M_{\Lambda_\alpha}(s)}(\cdot) =
\int_{-\infty}^{\infty} \chi_{\gamma}(t) \left(\int_0^t
\tau_u^{M_{\Lambda_\alpha}(s)}(\cdot) \,du \right)dt,
\end{equation}
with $M_{\Lambda_\alpha}(s) = \mathbb{I}_{C'}\otimes
H_{\Lambda_\alpha} + \frac{\Delta E}{4} V(s) +
\left(\frac{\mathbb{I}_{C'} + \sigma^z_{C'}}{2}\right)\otimes H_I$
and
\begin{equation}
H_{\Lambda_\alpha} = \sum_{{j}\in\Lambda_{\alpha}} h_{{j}},
\end{equation}
where $h_j = \mathcal{T}^{j}(h)$. Obviously $k_{\alpha}(s)$ has
support $\supp(k_{\alpha}(s)) = \Lambda_\alpha$.

We want to show that the quasi-adiabatic dynamics generated by
$k_{\alpha}(s)$:
\begin{equation}
\frac{d}{ds}\mathcal{V}_{\Lambda_\alpha}(s;0) =
ik_{\alpha}(s)\mathcal{V}_{\Lambda_\alpha}(s;0),
\end{equation}
are close to the quasi-adiabatic dynamics $\mathcal{V}(t;0)$
generated by $k(s)$. We do this by exploiting the inequality
\begin{equation*}
\|\mathcal{V}(t;0)-\mathcal{V}_{\Lambda_{\alpha}}(t;0)\| \le
\int_{0}^{|t|} \|k(s)-k_{{\alpha}}(s)\| ds,
\end{equation*}
which is proved, for example, by exploiting the Lie-Trotter
expansion.

We now show how the Lieb-Robinson bound provides an estimate on the
decay of $\|k_\alpha(s)\|$. Consider
\begin{multline}\label{eq:kbound1}
\begin{aligned}
&\|k(s)-k_{\alpha}(s)\| = \\
&\left\|\int_{-\infty}^{\infty} \chi_{\gamma}(t) \left(\int_0^t
\left(\tau_u^{M(s)}(m') -
\tau_u^{M_{\Lambda_{\alpha}}(s)}(m') \right)du \right)dt\right\| \\
&\le 2\int_{0}^{\infty} |\chi_{\gamma}(t)| \left(\int_0^t
\left\|\tau_u^{M(s)}(m') -
\tau_u^{{M_{\alpha}}(s)}(m') \right\|du\right) dt \\
&\le 2\int_{0}^{\infty} |\chi_{\gamma}(t)| \left(\int_0^t
\min\{2\|m'\|, c\alpha^2 \Delta
Ee^{\kappa |u| - v\alpha}\}du\right) dt \\
&\lesssim \frac{\alpha^2\Delta
E}{\gamma}\int_{0}^{c\alpha}|\chi_\gamma(t)|e^{\kappa |t| -
v\alpha}dt + \Delta
E\int_{c\alpha}^\infty |\chi_\gamma(t)||t|dt \\
&\lesssim \frac{\alpha^2\Delta E}{\gamma}\int_{0}^{c\alpha}
e^{\kappa t - v\alpha} dt + \Delta
E\int_{c\alpha}^\infty \frac{e^{-\frac{t^2}{2\gamma^2}}}{\sqrt{2\pi}\gamma} |t| dt \\
&\lesssim \frac{\alpha^2\Delta E}{\gamma}e^{(\kappa c - v)\alpha} +
{\gamma\Delta E}e^{-\frac{c^2\alpha^2}{2\gamma^2}},
\end{aligned}
\end{multline}
where $m' = \frac{\partial M(s)}{\partial s} = \frac{\Delta
E}{4}\frac{\partial V(s)}{\partial s}$, and in the first line we
applied the triangle inequality, in the second line we applied the
Lieb-Robinson bound, and in the third line we've broken the integral
into two pieces and applied the different regimes of the
Lieb-Robinson bound separately with $c$ some constant and we've used
the fact that $\|m'\|\lesssim \Delta E$. Thus, by choosing $c <
v/\kappa$ we see that $\|k(s) - k_\alpha(s)\|$ is decaying
exponentially fast in $\alpha$ for $\alpha \gtrsim \gamma$.

Thus we learn that the quasi-adiabatic dynamics
$\mathcal{V}(\frac{\pi}{2};0)$ are exponentially close (in operator
norm) to a unitary operator
$\mathcal{V}^{[1]}_{\Lambda_{\alpha}}(\frac{\pi}{2};0)$ which acts
nontrivially on only the region $\Lambda_\alpha$.

In order to complete our discussion and show that the state
$\mathcal{V}^{[1]}_{\Lambda_{\alpha}}(\frac{\pi}{2};0)|\Omega(0)\rangle$
is close to $|\Omega_H\rangle =
\mathcal{U}(\frac{\pi}{2};0)|\Omega(0)\rangle$ we apply the triangle
inequality to bound
\begin{multline}
\||\Omega_H\rangle -
\mathcal{V}^{[1]}_{\Lambda_{\alpha}}\left(\frac{\pi}{2};0\right)|\Omega(0)\rangle\|
\le \||\Omega_H\rangle -
\mathcal{V}\left(\frac{\pi}{2};0\right)|\Omega(0)\rangle\| +
\\ \|\mathcal{V}\left(\frac{\pi}{2};0\right)|\Omega(0)\rangle -
\mathcal{V}^{[1]}_{\Lambda_{\alpha}}\left(\frac{\pi}{2};0\right)|\Omega(0)\rangle
\|
\end{multline}
We use Eq.~(\ref{eq:error5}) from the appendix to bound the first
term and Eq.~(\ref{eq:kbound1}) to bound the second term:
\begin{equation}
\||\Omega_H\rangle -
\mathcal{V}^{[1]}_{\Lambda_{\alpha}}\left(\frac{\pi}{2};0\right)|\Omega(0)\rangle\|
\lesssim \frac{e^{-{2\gamma^2|x|^2\Delta E^2}}}{|x|\Delta E} +
\gamma\Delta E e^{-\frac{\kappa\gamma}{v}},
\end{equation}
so that $\gamma \gtrsim \frac{1}{|x|\Delta E}$ is sufficient to
ensure that the approximation is exponentially small (in $\gamma$).

With the choice $\alpha \gtrsim \gamma \gtrsim \frac{1}{|x|\Delta
E}$ we find that the width of the region $\Lambda_\alpha$ that the
unitary operator $\mathcal{V}^{[1]}_{\Lambda_{\alpha}}$ acts on is
given by $\frac{c}{|x|\Delta E}$, where $c$ is some constant.

To conclude our discussion we now show how to iteratively use the
procedure we've described in the previous paragraphs to construct an
approximation to the ground state $|\Omega_{H_n}\rangle$ of $H_n$
which can be stored (as a FCS) with resources scaling as a
polynomial in $n$.

Our first step is to start with a system  $A_1B_1$ which consists of
two copies $A_1$ and $A_2$ of the chain $\mathcal{C}_m$ on $m$ sites
with $m \gtrsim \frac{c}{|x|\Delta E}$. The hamiltonian for this
system is given by
\begin{equation}
K_m = H_m \otimes \mathbb{I}_{B_1} + \mathbb{I}_{A_1}\otimes H_m -
\delta_m\mathbb{I}.
\end{equation}
This system has a ground state equal to
$|\Omega_m\rangle|\Omega_m\rangle$. We then follow the procedure
described above, namely, adjoining an ancilla spin $C'$ between the
two blocks, and then applying the approximate adiabatic evolution
$\mathcal{V}^{[1]}_{\Lambda_{\alpha}}$ to yield an approximation
$\mathcal{V}^{[1]}_{\Lambda_{\alpha}}|\Omega_m\rangle|\Omega_m\rangle$
to the ground state $|\Omega_{2m}\rangle$.

We next iterate this procedure: we use two copies of the
approximation
$\mathcal{V}^{[1]}_{\Lambda_{\alpha}}|\Omega_m\rangle|\Omega_m\rangle$
to approximate the state $|\Omega_{2m}\rangle |\Omega_{2m}\rangle$.
This is the ground state of a new system $A_2B_2$ whose hamiltonian
is given by
\begin{equation}
K_{2m} = H_{2m} \otimes \mathbb{I}_{B_2} + \mathbb{I}_{A_2}\otimes
H_{2m} - \delta_{2m}\mathbb{I}.
\end{equation}
By the discussion above, we find that this hamiltonian is
adiabatically connected to $H_{4m}$. Because we are using an
approximation $|\Omega_{2m}'\rangle =
\mathcal{V}^{[1]}_{\Lambda_{\alpha}}|\Omega_m\rangle|\Omega_m\rangle$
for $|\Omega_{2m}\rangle$ we must account for the error that arises
when we take two copies of our approximation:
\begin{equation}
\begin{split}
\||\Omega_{2m}\rangle|\Omega_{2m}\rangle -
|\Omega_{2m}'\rangle|\Omega_{2m}'\rangle\| &\le
2\|(|\Omega_{2m}\rangle-|\Omega_{2m}'\rangle)|\Omega_{2m}\rangle\|
\\
&= 2\||\Omega_{2m}\rangle -
\mathcal{V}^{[1]}_{\Lambda_{\alpha}}|\Omega_m\rangle|\Omega_m\rangle\| \\
&\le 2\frac{e^{-{2\gamma^2|x|^2\Delta E^2}}}{|x|\Delta E} +
2\gamma\Delta E e^{-\frac{\kappa\gamma}{v}}
\end{split}
\end{equation}
Finally, using this upper bound we can compute the error between
$|\Omega_{4m}\rangle$ and our approximation (obtained from
approximating the adiabatic continuation from
$|\Omega_{2m}\rangle|\Omega_{2m}\rangle$ to $|\Omega_{4m}\rangle$):
\begin{equation}
\||\Omega_{4m}\rangle -
\mathcal{V}^{[2]}_{\Lambda_{\alpha}}|\Omega_{2m}'\rangle)|\Omega_{2m}'\rangle\|
\le 2\epsilon(\gamma) +\epsilon(\gamma),
\end{equation}
where we've defined
\begin{equation}
\epsilon(\gamma) = \frac{e^{-{2\gamma^2|x|^2\Delta E^2}}}{|x|\Delta
E} + \gamma\Delta E e^{-\frac{\kappa\gamma}{v}}.
\end{equation}
To obtain an approximation to the ground state $|\Omega_n\rangle$ of
$H_n$ we iterate the above procedure $\lceil \log_2(n/m)\rceil$
times. The error resulting from applying the above procedure is
given by
\begin{equation}
\frac{n}{m} \epsilon(\gamma) =
\frac{n}{m}\left(\frac{e^{-{2\gamma^2|x|^2\Delta E^2}}}{|x|\Delta E}
+ \gamma\Delta E e^{-\frac{\kappa\gamma}{v}}\right).
\end{equation}
If we fix some prespecified error $\epsilon$ and demand that our
approximation satisfies $\||\Omega_n\rangle -|\Omega_n'\rangle\|\le
\epsilon$ then we need that
\begin{equation} \gamma \ge
\max\left\{\frac{c}{|x|\Delta E},
c'\log\left(\frac{n}{\epsilon}\right)\right\},
\end{equation}
where $c$ and $c'$ are constants that only depend on $\|h\|$. Hence,
we learn that, similarly, the unitary operators
$\mathcal{V}^{[k]}_{\Lambda_\alpha}(\frac{\pi}{2})$ that we apply at
each stage $k$ act on a collection of $\alpha \ge c''\gamma$ spins,
with $c''$ some constant which only depends on $\|h\|$.

After applying the iterative procedure described above we end up
with the following representation for our approximation to
$|\Omega_n\rangle$:
\begin{equation}\label{eq:fcsrep1}
|\Omega_n\rangle = \mathcal{W}|\Omega_m\rangle|\Omega_m\rangle\cdots
|\Omega_m\rangle
\end{equation}
where
\begin{equation}
\mathcal{W} =
\mathcal{V}_{\Lambda_\alpha(1)}\mathcal{V}_{\Lambda_\alpha(2)}\cdots
\mathcal{V}_{\Lambda_\alpha(\lceil \log_2(n/m)\rceil-1)},
\end{equation}
with $\Lambda_\alpha(j) = \{k\,|\, d(k,mj) \le \alpha \}$.

The representation Eq.~(\ref{eq:fcsrep1}) is equivalent
\cite{osborne:2005d, verstraete:2005a} to a finitely correlated
state requiring a number of degrees of freedom which scale as
$n2^{c|\Lambda_{\alpha}|}$, where $c$ is some constant.
Alternatively, it is clear that the representation
Eq.~(\ref{eq:fcsrep1}) is already in a form useful for extracting
local properties: the expectation values of local operators such as
correlators are easy to compute using the representation
Eq.~(\ref{eq:fcsrep1}). Finally, it is worth noting that our
representation is also exactly in the form of a simple instance of
the \emph{multiscale entanglement renormalisation ansatz} introduced
in \cite{vidal:2005a}.

\section{Conclusions and future directions}
In this paper we have shown how a class of noncritical 1D quantum
spin systems are adiabatically connected to a 1D quantum spin system
of $\frac{n}{m}$ noninteracting quantum spins with local dimension
$2^{m}$. As long as $m \gg \max\left\{\frac{c}{|x|\Delta E},
c'\log\left(\frac{n}{\epsilon}\right)\right\}$, where $c$ and $c'$
are constants which only depend on $\|h\|$, then the ground state
$|\Omega_n\rangle$ of $H_n$ can be approximated efficiently. This
result bears a superficial resemblance to a naive application of
real-space renormalisation: in real-space renormalisation one argues
that blocks of $\frac{c}{\Delta E}$ spins should be effectively
noninteracting. (I.e., after $\log(\frac{c}{\Delta E})$
renormalisation group transformations we should be very close to a
trivial fixed point.) While this intuition is clear it seems to be
extremely challenging to put this intuition on a rigourous footing.

One question presents itself at this point: for noncritical spin
systems with a gap $\Delta E$ do the boundary effects persist only a
distance $\frac{c}{\Delta E}$ into the interior of the system?
Despite the plausibility of this statement we've been unable to
prove it; there are many counterexample systems which appear to ruin
the most obvious approaches.

\appendix
\section{Errors in approximations of adiabatic
dynamics}\label{app:appendix1}

The dynamics generated by the adiabatic evolution of a quantum
system is an important paradigm in the study of quantum mechanics.
One question which is particularly pertinent for the simulation of
quantum systems is: can adiabatic evolution be simulated by the
time-dependent dynamics of some (possibly different) quantum system?
In the case of adiabatic evolution the answer is yes: one method is
to just slowly turn on the interactions on a timescale which is
small compared to the gap of the system. While this is an entirely
satisfactory solution in and of itself one limitation lies in the
error of this approximation; when adiabatic dynamics is simulated
via this method the error decreases as an inverse polynomial in the
timescale $T$ of the slowly changing dynamics
\cite{reichardt:2004a}. Thus we are inspired to search for methods
which provide better error scaling. One such method is provided by
\emph{quasi-adiabatic evolution} \cite{wen:2005a} (see also
\cite{avron:1999a}). In this case we find that the error scales
\emph{exponentially} as a function of the timescale $T$ set by the
gap of the system. In this appendix we show this exponential
scaling. Our discussion is set in the wider context of ``adiabatic''
evolutions whose differential generators can be arbitrary.

The framework we describe in this section to discuss quasi-adiabatic
evolutions is similar to that introduced by \cite{avron:1999a} and
\cite{wen:2005a}.

\subsection{Quasi-adiabatic evolutions} In this subsection we introduce
a general framework to describe ``adiabatic''-like evolutions for
quantum systems.

We consider adiabatic quantum evolution generated by a
parameter-dependent hamiltonian $H(s)$ as $s$ is varied
adiabatically from $s=0$ to $s=1$. Thus we would like to understand
the ground state $|\Omega(s)\rangle$ of $H(s)$. We do this by
setting up a differential equation for $|\Omega(s)\rangle$:
\begin{equation}\label{eq:pformula}
\frac{d}{ds} |\Omega(s)\rangle = P'(s)|\Omega(s)\rangle,
\end{equation}
where $P'(s) = \frac{d}{ds}(|\Omega(s)\rangle\langle\Omega(s)|)$ and
we've set phases \cite{endnote30} so that $\langle
\Omega'(s)|\Omega(s)\rangle = 0$. Because $P'(s)$ is not
antihermitian the dynamics generated by this equation are not
unitary.

There are at least two ways to set up differential equations for
$|\Omega(s)\rangle$ which \emph{do} generate unitary dynamics. The
first is via \emph{exact adiabatic evolution} (see
\cite{avron:1987a, avron:1993a} for a rigourous discussion of rather
general results about exact adiabatic evolution):
\begin{equation}\label{eq:gpformula}
\frac{d}{ds} |\Omega(s)\rangle = -[P(s), P'(s)]|\Omega(s)\rangle.
\end{equation}
Because of the gap condition on $H(s)$, the ``hamiltonian''
$[P(s),P'(s)]$ for this dynamics is given by first-order stationary
perturbation theory:
\begin{multline}
[P(s), P'(s)] = |\Omega(s)\rangle \langle \Omega(s)| \frac{\partial
H(s)}{\partial s} \frac{\mathbb{I}}{\Omega(s)\mathbb{I}-H(s)} - \\
\frac{\mathbb{I}}{\Omega(s)\mathbb{I}-H(s)} \frac{\partial
H(s)}{\partial s} |\Omega(s)\rangle \langle \Omega(s)|,
\end{multline}
where $\Omega(s)$ is the ground-state energy of $H(s)$, and we
define $\frac{\mathbb{I}}{\Omega(s)\mathbb{I}-H(s)}$ via the
Moore-Penrose inverse:
$\frac{\mathbb{I}}{\Omega(s)\mathbb{I}-H(s)}|\Omega(s)\rangle =0$.

We now define an infinitesimal generator for a quantum evolution
which is meant to simulate the adiabatic evolution. We begin by
specifying a function $\chi_{\gamma}(t)$ which is an even real
function whose fourier transform $\widehat{\chi}_\gamma$ is
\emph{decaying rapidly} outside some region $[-\gamma, \gamma]$, and
which is normalised so that $\widehat{\chi}_\gamma(0) =1$. We use
this function to create an operator $Q(s)$ which is meant to
approximate $P(s)$:
\begin{equation}\label{eq:gsproj}
Q(s) = \int_{-\infty}^{\infty} \chi_{\gamma}(t)
e^{-it\Omega(s)}e^{itH(s)}dt,
\end{equation}
The following formula for $Q(s)$ may be verified by writing
$e^{itH}$ in its eigenbasis and exploiting the $L_2$ unitarity of
the fourier transform:
\begin{equation}
Q(s) = \sum_{j=0}^{2^n-1} \widehat{\chi}_\gamma(E_j(s) - \Omega(s))
|E_j(s)\rangle \langle E_j(s)|
\end{equation}

How close is $Q(s)$ to $P(s)$? We measure distance in operator norm:
\begin{equation}
\|Q(s)-P(s)\| = \sup_{z \in \spectrum(H(s))\setminus \Omega(s)}
\widehat{\chi}_\gamma(z - \Omega(s)).
\end{equation}
This formula shows us that, in the case where $H(s)$ has a gap
$\Delta E(s) \ge \Delta$, $Q(s)$ is close $P(s)$ in operator norm as
long as $|\widehat{\chi}_\gamma(z)|$ decays rapidly for $|z| \gtrsim
\Delta$.

In the case that $\chi_{\gamma}(t)$ is an even real function whose
fourier transform $\widehat{\chi}_\gamma$ is $C^\infty$, has compact
support in $[-\gamma, \gamma]$, and is normalised so that
$\widehat{\chi}_\gamma(0) =1$ with $\gamma < \Delta$ to ensure that
only the ground state appears on the RHS of (\ref{eq:gsproj}) we can
recover exact adiabatic dynamics: we first use the Duhamel formula
\begin{equation*}
\frac{d}{ds}e^{itH(s)} =  i\int_0^t e^{i(t-u)H(s)}\frac{\partial
H(s)}{\partial s} e^{iuH(s)} du,
\end{equation*}
to rewrite (\ref{eq:pformula}):
\begin{widetext}
\begin{equation}
\frac{d}{ds} |\Omega(s)\rangle = -i \frac{d\Omega(s)}{ds}
\int_{-\infty}^{\infty} t\chi_\gamma(t) dt|\Omega(s)\rangle +
i\int_{-\infty}^{\infty}
\chi_{\gamma}(t)e^{-it\Omega(s)}\left(\int_0^t
\tau_u^{H(s)}\left(\frac{\partial H(s)}{\partial s}\right) du\right)
 e^{itH(s)}dt|\Omega(s)\rangle.
\end{equation}
\end{widetext}
Using the fact that $\chi_\gamma(t)$ is an even function of $t$ and
cancelling phases we obtain
\begin{equation}\label{eq:exadinfgen}
\frac{d}{ds} |\Omega(s)\rangle = i\int_{-\infty}^{\infty}
\chi_{\gamma}(t)\left(\int_0^t \tau_u^{H(s)}\left(\frac{\partial
H(s)}{\partial s}\right) du\right)
 dt|\Omega(s)\rangle.
\end{equation}
By integrating this expression for $\frac{d}{ds} |\Omega(s)\rangle$
in the energy eigenbasis of $H(s)$ and using the assumed gap
structure one can find that this expression is equivalent to the
usual expression obtained from first-order perturbation theory:
\begin{equation}
\frac{d}{ds} |\Omega(s)\rangle =
\frac{\mathbb{I}}{\Omega(s)\mathbb{I} - H(s)}\frac{\partial
H(s)}{\partial s} |\Omega(s)\rangle.
\end{equation}

We use the form (\ref{eq:exadinfgen}) to deduce a \emph{hermitian}
infinitesimal generator
\begin{equation}\label{eq:exadinfinfgen}
K(s) = \int_{-\infty}^{\infty} \chi_{\gamma}(t)\left(\int_0^t
\tau_u^{H(s)}\left(\frac{\partial H(s)}{\partial s}\right) du\right)
dt
\end{equation}
of exact adiabatic evolution.  Thus we find
\begin{equation}
|\Omega(s)\rangle = \mathcal{T} e^{i\int^{t}_0 K(s) \,ds}
|\Omega(0)\rangle = \mathcal{U}(t;0)|\Omega(0)\rangle
\end{equation}
where we define
\begin{equation}\label{eq:adevol}
\mathcal{U}(t;0) = \mathcal{T} e^{i\int^{t}_0 K(s) \,ds}.
\end{equation}
It is now easy to guess a form for the infinitesimal generator
$L_{\chi_\gamma}(s)$ which is meant to approximate exact adiabatic
dynamics: we just allow the function $\chi_\gamma(s)$ in
(\ref{eq:exadinfinfgen}) to be arbitrary:
\begin{equation}\label{eq:exadinfinfgen2}
L_{\chi_\gamma}(s) = \int_{-\infty}^{\infty}
\chi_{\gamma}(t)\left(\int_0^t \tau_u^{H(s)}\left(\frac{\partial
H(s)}{\partial s}\right) du\right) dt.
\end{equation}
(We drop the subscript on $L_{\chi_\gamma}(s)$ from now on.) The
corresponding dynamics are obtained by integration:
\begin{equation}\label{eq:qadevol}
\mathcal{V}(t;0) = \mathcal{T} e^{i\int^{t}_0 L(s) \,ds}.
\end{equation}
We call the dynamics $\mathcal{V}(t;0)$ a \emph{quasi-adiabatic
evolution}.

With the definition (\ref{eq:qadevol}) of quasi-adiabatic evolution
we now define the (time-dependent) state $|\Phi(t)\rangle$:
\begin{equation}
|\Phi(t)\rangle = \mathcal{V}(t;0)|\Omega(0)\rangle.
\end{equation}
The idea is that $|\Phi(t)\rangle$ should be very close to
$|\Omega(t)\rangle$ as long as $\|Q(s)-P(s)\|$ is small. We make
this rigourous in the next section.

\subsection{Error in quasi-adiabatic evolution} In this section we
study the error $\delta(t) = \| |\Phi(t)\rangle - |\Omega(t)\rangle
\|$ between the exact ground state of $H(s)$ and the state
$|\Phi(t)\rangle$ generated by quasi-adiabatic evolution.

To get started on an upper bound for $\delta(t)$ we define
\begin{equation}
|\delta(t)\rangle = |\Omega(0)\rangle -
\mathcal{V}(t;0)^\dag|\Omega(t)\rangle = |\Omega(0)\rangle -
\mathcal{V}(0;t)|\Omega(t)\rangle.
\end{equation}
It is easy to see, using the unitary invariance of $\|\cdot\|$, that
$\delta(t) = \| |\delta(t)\rangle \|$.

We use the fundamental theorem of calculus to write
$|\delta(t)\rangle$ as
\begin{equation}
|\delta(t)\rangle = \int_0^{t} \frac{d}{ds}
\left(\mathcal{V}(s;0)^\dag|\Omega(s)\rangle\right)\, ds.
\end{equation}
Now we use the definitions of $\frac{d}{ds}|\Omega(s)\rangle$ and
$\frac{d}{ds} \mathcal{V}(s;0)$ to find
\begin{widetext}
\begin{equation}
|\delta(t)\rangle = \int_0^{t} \left(-i\mathcal{V}(s;0)^\dag L(s)
|\Omega(s)\rangle + \mathcal{V}(s;0)^\dag
\frac{\mathbb{I}}{\Omega(s)\mathbb{I} - H(s)}\frac{\partial
H(s)}{\partial s}|\Omega(s)\rangle\right)\, ds.
\end{equation}
\end{widetext} Thus, using the triangle inequality and unitary
invariance of $\|\cdot\|$, we find
\begin{multline}\label{eq:error1}
\delta(t) = \| |\delta(t)\rangle \| \le\\ \int_0^{t} \left\| iL(s)
|\Omega(s)\rangle - \frac{\mathbb{I}}{\Omega(s)\mathbb{I} -
H(s)}\frac{\partial H(s)}{\partial s}|\Omega(s)\rangle\right\|\, ds.
\end{multline}
To make some headway on this expression we derive an expression for
$L(s) |\Omega(s)\rangle$:
\begin{multline}
iL(s) |\Omega(s)\rangle = \\ i\int_{-\infty}^{\infty}
\chi_{\gamma}(t)\left(\int_0^t \tau_u^{H(s)}\left(\frac{\partial
H(s)}{\partial s}\right) du\right) dt |\Omega(s)\rangle.
\end{multline}
By integration, we find
\begin{multline}
iL(s) |\Omega(s)\rangle = \\
\left(\frac{\widehat{\chi}_\gamma(H(s)-\Omega(s)\mathbb{I})}{H(s)-\Omega(s)\mathbb{I}}
 + \frac{\mathbb{I}}{\Omega(s)\mathbb{I}-H(s)}\right)\frac{\partial H(s)}{\partial s}
 |\Omega(s)\rangle,
\end{multline}
where we define, as before,
$\frac{\mathbb{I}}{\Omega(s)\mathbb{I}-H(s)}$ via the Moore-Penrose
inverse:
$\frac{\mathbb{I}}{\Omega(s)\mathbb{I}-H(s)}|\Omega(s)\rangle =0$,
and we define the operator
$\widehat{\chi}_\gamma(H(s)-\Omega(s)\mathbb{I})$ using the
holomorphic function calculus \cite{kadison:1997a}:
\begin{equation}
\widehat{\chi}_\gamma(H(s)-\Omega(s)\mathbb{I}) = \sum_{j=0}^{2^n -
1} \widehat{\chi}_\gamma(E_j(s)-\Omega(s)) |E_j(s)\rangle \langle
E_j(s)|.
\end{equation}
Substituting this expression for $iL(s) |\Omega(s)\rangle$ into
(\ref{eq:error1}) we find
\begin{equation}\label{eq:error2}
\delta(t) \le \int_0^{t} \left\|
\frac{\widehat{\chi}_\gamma(H(s)-\Omega(s)\mathbb{I})}{H(s)-\Omega(s)\mathbb{I}}\frac{\partial
H(s)}{\partial s}|\Omega(s)\rangle\right\|\, ds.
\end{equation}
We rewrite this as
\begin{equation}\label{eq:error3}
\delta(t)\le \int_0^{t} \sqrt{\langle
\eta(s)|\left(\frac{\widehat{\chi}_\gamma(H(s)-\Omega(s)\mathbb{I})}{H(s)-\Omega(s)\mathbb{I}}\right)^2
|\eta(s)\rangle} \, ds,
\end{equation}
where
\begin{equation}
|\eta(s)\rangle = \mathcal{P}_{\text{high}} \frac{\partial
H(s)}{\partial s}|\Omega(s)\rangle
\end{equation}
and
\begin{equation}
\mathcal{P}_{\text{high}} = \sum_{j =1}^{2^n-1} |E_j(s)\rangle
\langle E_j(s)|.
\end{equation}
(Recall we are defining
$\frac{\mathbb{I}}{\Omega(s)\mathbb{I}-H(s)}$ via the Moore-Penrose
inverse.)

We now define $\eta(t) =
\left\|\mathcal{P}_{\text{high}}\frac{\partial H(s)}{\partial
s}|\Omega(s)\rangle\right\| \le \left\|\frac{\partial H(s)}{\partial
s}\right\|$ and $\eta_* = \sup_{s\in[0, t]} \eta(s)$ to rewrite
(\ref{eq:error3}) as
\begin{equation}\label{eq:error4}
\| |\delta(t)\rangle \| \le \eta_*\int_0^{t}
\left\|\frac{\widehat{\chi}_\gamma(H(s)-\Omega(s)\mathbb{I})}{H(s)-\Omega(s)\mathbb{I}}\right\|
\, ds.
\end{equation}
We also define
\begin{equation}
\begin{split}
f(s) &=
\left\|\frac{\widehat{\chi}_\gamma(H(s)-\Omega(s)\mathbb{I})}{H(s)-\Omega(s)\mathbb{I}}\right\|
\\
&= \sup_{z \in \spectrum(H(s))\setminus \Omega(s)}
\frac{\widehat{\chi}_\gamma(z-\Omega(s))}{z-\Omega(s)}
\end{split}
\end{equation}
and set $f_* = \sup_{s\in[0, t]} f(s)$ to obtain our final estimate
\begin{equation}\label{eq:error5}
\| |\delta(t)\rangle \| \le \eta_*f_*.
\end{equation}

The quantities $\eta_*$ and $f_*$ can be separately upper bounded.
In the case of spin systems which are adiabatically evolving we can
generally bound $\eta_*$ by $\eta_* \le cn$, where $c$ is a constant
and $n$ is the number of spins. For $f_*$: in the case that our
system has a gap $\Delta$ we use the cutoff function
\begin{equation}
\chi_\gamma(t) =
\frac{e^{-\frac{t^2}{2\gamma^2}}}{\sqrt{2\pi}\gamma},
\end{equation}
which has fourier transform
\begin{equation}
\widehat{\chi}_\gamma(\omega) = {e^{-{2\gamma^2\omega^2}}}.
\end{equation}
We can now put together the gap structure of the spectrum of $H(s)$
and the fourier transform of $\chi_\gamma(t)$ to obtain the bound
\begin{equation}
f_* \le \frac{e^{-{2\gamma^2\Delta^2}}}{\Delta}.
\end{equation}
This expression is exponentially decaying in $\gamma$, indeed,
$\gamma \gtrsim \frac{1}{\Delta}$ is sufficient to make this upper
bound for $f_*$ arbitrarily small. Putting these two bounds for
$\eta_*$ and $f_*$ together we find that, for adiabatically evolving
spin systems, $\gamma \gtrsim \frac{1}{\Delta}\sqrt{\log(n)}$ is
sufficient to reduce the error $\delta(t)$ until it is arbitrarily
small.

It is easy to extend our discussion to systems $H(s)$ which have a
continuous gapless spectrum as long as there is some bound on the
growth of the density of states near the ground state
\cite{avron:1999a, wen:2005a}.


\begin{thebibliography}{27}
\expandafter\ifx\csname
natexlab\endcsname\relax\def\natexlab#1{#1}\fi
\expandafter\ifx\csname bibnamefont\endcsname\relax
  \def\bibnamefont#1{#1}\fi
\expandafter\ifx\csname bibfnamefont\endcsname\relax
  \def\bibfnamefont#1{#1}\fi
\expandafter\ifx\csname citenamefont\endcsname\relax
  \def\citenamefont#1{#1}\fi
\expandafter\ifx\csname url\endcsname\relax
  \def\url#1{\texttt{#1}}\fi
\expandafter\ifx\csname urlprefix\endcsname\relax\def\urlprefix{URL
}\fi \providecommand{\bibinfo}[2]{#2}
\providecommand{\eprint}[2][]{\url{#2}}

\bibitem[{\citenamefont{Hastings}(2005)}]{hastings:2006a}
\bibinfo{author}{\bibfnamefont{M.~B.} \bibnamefont{Hastings}},
  \bibinfo{journal}{Phys. Rev. B} \textbf{\bibinfo{volume}{73}},
  \bibinfo{pages}{085115} (\bibinfo{year}{2005}), \eprint{cond-mat/0508554}.

\bibitem[{\citenamefont{Schollw{\"o}ck}(2005)}]{schollwoeck:2005a}
\bibinfo{author}{\bibfnamefont{U.}~\bibnamefont{Schollw{\"o}ck}},
  \bibinfo{journal}{Rev. Modern Phys.} \textbf{\bibinfo{volume}{77}},
  \bibinfo{pages}{259} (\bibinfo{year}{2005}), \eprint{cond-mat/0409292}.

\bibitem[{\citenamefont{Verstraete and Cirac}(2004)}]{verstraete:2004a}
\bibinfo{author}{\bibfnamefont{F.}~\bibnamefont{Verstraete}} \bibnamefont{and}
  \bibinfo{author}{\bibfnamefont{J.~I.} \bibnamefont{Cirac}}
  (\bibinfo{year}{2004}), \eprint{cond-mat/0407066}.

\bibitem[{\citenamefont{Eisert}(2006)}]{eisert:2006a}
\bibinfo{author}{\bibfnamefont{J.}~\bibnamefont{Eisert}},
  \bibinfo{journal}{Phys. Rev. Lett.} \textbf{\bibinfo{volume}{62}},
  \bibinfo{pages}{260501} (\bibinfo{year}{2006}), \eprint{quant-ph/0609051}.

\bibitem[{\citenamefont{Fannes et~al.}(1992)\citenamefont{Fannes, Nachtergaele,
  and Werner}}]{fannes:1992a}
\bibinfo{author}{\bibfnamefont{M.}~\bibnamefont{Fannes}},
  \bibinfo{author}{\bibfnamefont{B.}~\bibnamefont{Nachtergaele}},
  \bibnamefont{and} \bibinfo{author}{\bibfnamefont{R.~F.}
  \bibnamefont{Werner}}, \bibinfo{journal}{Comm. Math. Phys.}
  \textbf{\bibinfo{volume}{144}}, \bibinfo{pages}{443} (\bibinfo{year}{1992}).

\bibitem[{\citenamefont{Vidal}(2003)}]{vidal:2003a}
\bibinfo{author}{\bibfnamefont{G.}~\bibnamefont{Vidal}},
  \bibinfo{journal}{Phys. Rev. Lett.} \textbf{\bibinfo{volume}{93}},
  \bibinfo{pages}{040502} (\bibinfo{year}{2003}), \eprint{quant-ph/0310089}.

\bibitem[{\citenamefont{Zwolak and Vidal}(2003)}]{zwolak:2004a}
\bibinfo{author}{\bibfnamefont{M.}~\bibnamefont{Zwolak}} \bibnamefont{and}
  \bibinfo{author}{\bibfnamefont{G.}~\bibnamefont{Vidal}},
  \bibinfo{journal}{Phys. Rev. Lett.} \textbf{\bibinfo{volume}{93}},
  \bibinfo{pages}{207205} (\bibinfo{year}{2003}), \eprint{cond-mat/0406440}.

\bibitem[{\citenamefont{Verstraete et~al.}(2004)\citenamefont{Verstraete,
  Garc{\'\i}a-Ripoll, and Cirac}}]{verstraete:2004b}
\bibinfo{author}{\bibfnamefont{F.}~\bibnamefont{Verstraete}},
  \bibinfo{author}{\bibfnamefont{J.~J.} \bibnamefont{Garc{\'\i}a-Ripoll}},
  \bibnamefont{and} \bibinfo{author}{\bibfnamefont{J.~I.} \bibnamefont{Cirac}},
  \bibinfo{journal}{Phys. Rev. Lett.} \textbf{\bibinfo{volume}{93}},
  \bibinfo{pages}{207204} (\bibinfo{year}{2004}), \eprint{quant-ph/0406426}.

\bibitem[{\citenamefont{Osborne}(2006{\natexlab{a}})}]{osborne:2006a}
\bibinfo{author}{\bibfnamefont{T.~J.} \bibnamefont{Osborne}}
  (\bibinfo{year}{2006}{\natexlab{a}}), \eprint{quant-ph/0601019}.

\bibitem[{\citenamefont{Oliveira and Terhal}(2005)}]{oliveira:2005a}
\bibinfo{author}{\bibfnamefont{R.}~\bibnamefont{Oliveira}} \bibnamefont{and}
  \bibinfo{author}{\bibfnamefont{B.~M.} \bibnamefont{Terhal}}
  (\bibinfo{year}{2005}), \eprint{quant-ph/0504050}.

\bibitem[{\citenamefont{Kempe et~al.}(2004)\citenamefont{Kempe, Kitaev, and
  Regev}}]{kempe:2004a}
\bibinfo{author}{\bibfnamefont{J.}~\bibnamefont{Kempe}},
  \bibinfo{author}{\bibfnamefont{A.}~\bibnamefont{Kitaev}}, \bibnamefont{and}
  \bibinfo{author}{\bibfnamefont{O.}~\bibnamefont{Regev}}, in
  \emph{\bibinfo{booktitle}{FSTTCS 2004: Foundations of software technology and
  theoretical computer science}} (\bibinfo{publisher}{Springer},
  \bibinfo{address}{Berlin}, \bibinfo{year}{2004}), vol. \bibinfo{volume}{3328}
  of \emph{\bibinfo{series}{Lecture Notes in Comput. Sci.}}, pp.
  \bibinfo{pages}{372--383}, \eprint{quant-ph/0406180}.

\bibitem[{\citenamefont{Kitaev et~al.}(2002)\citenamefont{Kitaev, Shen, and
  Vyalyi}}]{kitaev:2002a}
\bibinfo{author}{\bibfnamefont{A.~Y.} \bibnamefont{Kitaev}},
  \bibinfo{author}{\bibfnamefont{A.~H.} \bibnamefont{Shen}}, \bibnamefont{and}
  \bibinfo{author}{\bibfnamefont{M.~N.} \bibnamefont{Vyalyi}},
  \emph{\bibinfo{title}{Classical and quantum computation}},
  vol.~\bibinfo{volume}{47} of \emph{\bibinfo{series}{Graduate Studies in
  Mathematics}} (\bibinfo{publisher}{American Mathematical Society},
  \bibinfo{address}{Providence, RI}, \bibinfo{year}{2002}).

\bibitem[{\citenamefont{Bhatia}(1997)}]{bhatia:1997a}
\bibinfo{author}{\bibfnamefont{R.}~\bibnamefont{Bhatia}},
  \emph{\bibinfo{title}{Matrix analysis}}
  (\bibinfo{publisher}{Springer-Verlag}, \bibinfo{address}{New York},
  \bibinfo{year}{1997}).

\bibitem[{\citenamefont{Lieb and Robinson}(1972)}]{lieb:1972a}
\bibinfo{author}{\bibfnamefont{E.~H.} \bibnamefont{Lieb}} \bibnamefont{and}
  \bibinfo{author}{\bibfnamefont{D.~W.} \bibnamefont{Robinson}},
  \bibinfo{journal}{Commun. math. Phys.} \textbf{\bibinfo{volume}{28}},
  \bibinfo{pages}{251} (\bibinfo{year}{1972}).

\bibitem[{\citenamefont{Hastings}(2004)}]{hastings:2004a}
\bibinfo{author}{\bibfnamefont{M.~B.} \bibnamefont{Hastings}},
  \bibinfo{journal}{Phys. Rev. B} \textbf{\bibinfo{volume}{69}},
  \bibinfo{pages}{104431} (\bibinfo{year}{2004}), \eprint{cond-mat/0305505}.

\bibitem[{\citenamefont{Nachtergaele and Sims}(2006)}]{nachtergaele:2005a}
\bibinfo{author}{\bibfnamefont{B.}~\bibnamefont{Nachtergaele}}
  \bibnamefont{and} \bibinfo{author}{\bibfnamefont{R.}~\bibnamefont{Sims}},
  \bibinfo{journal}{Commun. Math. Phys.} \textbf{\bibinfo{volume}{265}},
  \bibinfo{pages}{119} (\bibinfo{year}{2006}), \eprint{math-ph/0506030}.

\bibitem[{\citenamefont{Hastings and Koma}(2006)}]{hastings:2005b}
\bibinfo{author}{\bibfnamefont{M.~B.} \bibnamefont{Hastings}} \bibnamefont{and}
  \bibinfo{author}{\bibfnamefont{T.}~\bibnamefont{Koma}},
  \bibinfo{journal}{Commun. Math. Phys.} \textbf{\bibinfo{volume}{265}},
  \bibinfo{pages}{781} (\bibinfo{year}{2006}), \eprint{math-ph/0507008}.

\bibitem[{\citenamefont{Osborne}(2006{\natexlab{b}})}]{osborne:2006b}
\bibinfo{author}{\bibfnamefont{T.~J.} \bibnamefont{Osborne}}
  (\bibinfo{year}{2006}{\natexlab{b}}),
  \bibinfo{note}{\texttt{www.lri.fr/qip06/slides/os\-borne.pdf}}.

\bibitem[{\citenamefont{Osborne}(2006{\natexlab{c}})}]{osborne:2005d}
\bibinfo{author}{\bibfnamefont{T.~J.} \bibnamefont{Osborne}},
  \bibinfo{journal}{Phys. Rev. Lett.} \textbf{\bibinfo{volume}{97}},
  \bibinfo{pages}{157202} (\bibinfo{year}{2006}{\natexlab{c}}),
  \eprint{quant-ph/0508031}.

\bibitem[{\citenamefont{Verstraete and Cirac}(2006)}]{verstraete:2005a}
\bibinfo{author}{\bibfnamefont{F.}~\bibnamefont{Verstraete}} \bibnamefont{and}
  \bibinfo{author}{\bibfnamefont{J.~I.} \bibnamefont{Cirac}},
  \bibinfo{journal}{Phys. Rev. B} \textbf{\bibinfo{volume}{73}},
  \bibinfo{pages}{094423} (\bibinfo{year}{2006}), \eprint{cond-mat/0505140}.

\bibitem[{\citenamefont{Vidal}(2005)}]{vidal:2005a}
\bibinfo{author}{\bibfnamefont{G.}~\bibnamefont{Vidal}} (\bibinfo{year}{2005}),
  \eprint{cond-mat/0512165}.

\bibitem[{\citenamefont{Reichardt}(2004)}]{reichardt:2004a}
\bibinfo{author}{\bibfnamefont{B.~W.} \bibnamefont{Reichardt}}, in
  \emph{\bibinfo{booktitle}{Proceedings of the 36th Annual ACM Symposium on
  Theory of Computing}} (\bibinfo{publisher}{ACM}, \bibinfo{address}{New York},
  \bibinfo{year}{2004}), pp. \bibinfo{pages}{502--510}.

\bibitem[{\citenamefont{Wen and Hastings}(2005)}]{wen:2005a}
\bibinfo{author}{\bibfnamefont{X.-G.} \bibnamefont{Wen}} \bibnamefont{and}
  \bibinfo{author}{\bibfnamefont{M.~B.} \bibnamefont{Hastings}},
  \bibinfo{journal}{Phys. Rev. B} \textbf{\bibinfo{volume}{72}},
  \bibinfo{pages}{045141} (\bibinfo{year}{2005}), \eprint{cond-mat/0503554}.

\bibitem[{\citenamefont{Avron and Elgart}(1999)}]{avron:1999a}
\bibinfo{author}{\bibfnamefont{J.~E.} \bibnamefont{Avron}} \bibnamefont{and}
  \bibinfo{author}{\bibfnamefont{A.}~\bibnamefont{Elgart}},
  \bibinfo{journal}{Comm. Math. Phys.} \textbf{\bibinfo{volume}{203}},
  \bibinfo{pages}{445} (\bibinfo{year}{1999}).

\bibitem[{\citenamefont{Avron et~al.}(1987)\citenamefont{Avron, Seiler, and
  Yaffe}}]{avron:1987a}
\bibinfo{author}{\bibfnamefont{J.~E.} \bibnamefont{Avron}},
  \bibinfo{author}{\bibfnamefont{R.}~\bibnamefont{Seiler}}, \bibnamefont{and}
  \bibinfo{author}{\bibfnamefont{L.~G.} \bibnamefont{Yaffe}},
  \bibinfo{journal}{Comm. Math. Phys.} \textbf{\bibinfo{volume}{110}},
  \bibinfo{pages}{33} (\bibinfo{year}{1987}).

\bibitem[{\citenamefont{Avron et~al.}(1993)\citenamefont{Avron, Seiler, and
  Yaffe}}]{avron:1993a}
\bibinfo{author}{\bibfnamefont{J.~E.} \bibnamefont{Avron}},
  \bibinfo{author}{\bibfnamefont{R.}~\bibnamefont{Seiler}}, \bibnamefont{and}
  \bibinfo{author}{\bibfnamefont{L.~G.} \bibnamefont{Yaffe}},
  \bibinfo{journal}{Comm. Math. Phys.} \textbf{\bibinfo{volume}{156}},
  \bibinfo{pages}{649} (\bibinfo{year}{1993}).

\bibitem[{\citenamefont{Kadison and Ringrose}(1997)}]{kadison:1997a}
\bibinfo{author}{\bibfnamefont{R.~V.} \bibnamefont{Kadison}} \bibnamefont{and}
  \bibinfo{author}{\bibfnamefont{J.~R.} \bibnamefont{Ringrose}},
  \emph{\bibinfo{title}{Fundamentals of the theory of operator algebras. {V}ol.
  {I}}}, vol.~\bibinfo{volume}{15} of \emph{\bibinfo{series}{Graduate Studies
  in Mathematics}} (\bibinfo{publisher}{American Mathematical Society},
  \bibinfo{address}{Providence, RI}, \bibinfo{year}{1997}).

\bibitem{endnote28}{By ``noncritical'' we mean the the Hamiltonian
for the system has a spectral gap between the ground state and
first-excited state which is a constant that doesn't scale with $n$,
the number of spins.}

\bibitem{endnote29}{Barbara Terhal and David DiVincenzo (private
communication)}

\bibitem{endnote30}{Note that this choice precludes an extension of our
analysis to study Berry phases.}


\end{thebibliography}
\end{document}